\documentclass[12pt]{article}
\usepackage{graphics}

\begin{document}

\title{Alternative Approach to
$B^- \to \eta^{\prime} K^-$ Branching Ratio Calculation}
\author{Marina-Aura Dariescu
and Ciprian Dariescu \\
Dept. of Theoretical Physics \\ {\it Al. I. Cuza} University \\
Bd. Carol I no. 11,
6600 Ia\c{s}i, Romania \\
email: marina@uaic.ro}
\date{}

\maketitle

\begin{flushleft}
Short title: \\
Alternative Approach to $B^- \to \eta^{\prime} K^-$ BR Calculation
\end{flushleft}

\begin{abstract}
Since the calculation of $BR(B^- \to \eta^{\prime} K^-)$ in the
framework of QCD improved factorization method, developed by
Beneke {\it et al.}, leads to numerical values much below the
experimental data, we include two different contributions, in an
alternative way. First, we find out that the spectator
hard-scattering mechanism increases the $BR$ value with almost $50
\%$, but the predictions depend on the combined singularities in
the amplitude convolution. Secondly, by adding SUSY contributions
to the Wilson coefficients, we come to a $BR$ depending on three
parameters, whose values are constrained by the experimental data.
\end{abstract}

\baselineskip 1.6em

\section{Introduction}

As a first evidence of a strong penguin, the $B^- \to
\eta^{\prime} K^-$ decay has become of a real interest after CLEO
announced its large numerical value $BR(B^- \to \eta^{\prime}K^- )
= ( 6.5^{+1.5}_{-1.4} \pm 0.9 ) \times 10^{-5}$ [1], which could
not be explained by the existent theoretical models. As improved
measurements followed, providing even larger values,
$(80^{+10}_{-9} \pm 7) \times 10^{-6}$ (CLEO [2]), $(76.9 \pm 3.5
\pm 4.4) \times 10^{-6}$ (BaBar [3]) and $(79^{+12}_{-11} \pm 9)
\times 10^{-6}$ (Belle [4]), the inclusion of new contributions
for accommodating these data has quickly become a real theoretical
challenge. In this respect, perturbative QCD mechanisms [5], with
different $\eta^{\prime} g^* g^*$ vertex function [5, 6], have
been considered as main candidates for significantly increasing
the $BR(B^- \to \eta^{\prime} K^-)$ value. On the other hand,
while searching for physics beyond the Standard Model (SM),
supersymmetry has been employed in processes like $B \to J/\psi
K^{*}$ [7], $B \to \phi K$ [8], $B \to \pi K$ [9, 10], $B \to X_s
\gamma$ [11], and deviations from the SM predictions for the
values of branching ratios and $CP$ asymmetries have been the main
targets.

The present paper is organized as follows: in Section 2, we
compute the $BR(B^- \to \eta^{\prime} K^-)$ in the improved
factorization approach developed by Beneke {\it et al.} [12].
Since we get a $BR$ much below the experimental values, we
incorporate two alternative contributions. The first one,
presented in Section 3, comes from the so-called spectator hard
scattering mechanism. Following a similar approach as in [13], we
give a detailed calculation of the gluonic transition form factor
which plays an important role in the evaluation of this
contribution. Although it has been concluded that this mechanism
could provide large $BR$ values [13], we show that the presence of
combined singularities in the amplitude convolution is a source of
large uncertainties. In Section 4, we employ a supersymmetric
approach and include exchanges of gluino and squark with
left-right squark mixing. Working in the mass insertion
approximation [14], the values of the Wilson coefficients $c_{8g}$
and $c_{7 \gamma}$ can be significantly increased, by adding the
SUSY contributions, and this has a strong numerical impact in the
branching ratio estimation. Finally, one may use the experimental
data to impose constraints on the flavor changing SUSY parameter
$\delta_{LR}^{bs}$.

\section{Improved QCD Factorisation}

The relevant decay amplitude
for $ B^- \to \eta^{\prime} K^-$, in the improved
QCD factorization approach [12], is given by [5, 15]
\begin{eqnarray}
A(B^- \to \eta^{\prime} K^-)
&=& - \, i {G_F \over \sqrt{2}}
(m^2_B - m^2_{\eta^{\prime}})
F^{B\to \eta^{\prime}}_0(m^2_K)
f_K [ V_{ub} V_{us}^* a_1(X)
\nonumber \\*
&& + \, V_{pb}V_{ps}^* \left( a^p_4(X)+ a^p_{10}(X)
+ r^K_\chi (a_6^p(X) + a_8^p(X)) \right) ]
\nonumber \\*
&-& i {G_F \over \sqrt{2}} (m^2_B - m^2_K)
F^{B\to K}_0(m^2_{\eta^{\prime}}) f^u_{\eta^{\prime}}
\left[ V_{ub}V_{us}^* a_2(Y) \right.
\nonumber \\*
&&
+ V_{pb}V_{ps}^* [ \left( a_3(Y)-a_5(Y) \right)
( 2 + \sigma )
\nonumber \\*
& & +
\left[ a_4^p (Y) - \frac{1}{2} a_{10}^p (Y)
+ r_{\chi}^{\prime} \left(
a_6^p(Y) - \frac{1}{2} a_8^p (Y) \right) \right]
\sigma
\nonumber \\*
& &
\left. + \frac{1}{2}
\left( a_9(Y) - a_7(Y) \right) (1 - \sigma ) \right] ,
\end{eqnarray}
where $X = \eta^{\prime} K$ and $Y = K \eta^{\prime}$,
$p$ is summed over $u$ and $c$, $r^{\prime}_\chi =
2m^2_{\eta^{\prime}}/(m_b-m_s)(2m_s)$,
$r^K_\chi = 2m^2_K/m_b(m_u + m_s)$,
$\sigma = f^s_{\eta^{\prime}}/f^u_{\eta^{\prime}}$,
and [12]
\begin{eqnarray}
a_1(M_1 M_2) & = & c_1 + {c_2 \over N_c}
\left[ 1 + {C_F \alpha_s \over 4 \pi}
(V_{M_2} +  H) \right],
\nonumber\\
a_2(M_1 M_2) & = & c_2 + {c_1 \over N_c}
\left[ 1 + {C_F \alpha_s \over 4 \pi}
(V_{M_2} +  H) \right],
\nonumber\\
a_3(M_1 M_2) & = & c_3 + {c_4 \over N_c}
\left[ 1 + {C_F \alpha_s \over 4 \pi}
(V_{M_2} +  H) \right],
\nonumber\\
a^p_4(M_1 M_2) & = &
c_4 + {c_3\over N_c} \left[ 1 + {C_F \alpha_s \over 4 \pi}
(V_{M_2} + H) \right] +
{C_F \alpha_s \over 4\pi N_c} P^p_{M_2,2},
\nonumber\\
a_5(M_1 M_2) & = & c_5 + {c_6 \over N_c}
\left[ 1 + {C_F \alpha_s \over 4 \pi}
(-12-V_{M_2} - H) \right],
\nonumber\\
a^p_6(M_1 M_2) & = & c_6 + {c_5\over N_c}
\left( 1 - 6 {C_F \alpha_s \over 4
\pi} \right) + {C_F\alpha_s \over 4\pi N_c}
P^{p}_{M_2, 3},
\nonumber\\
a_7(M_1 M_2) & = & c_7 + {c_8 \over N_c}
\left[ 1 + {C_F \alpha_s \over 4 \pi}
(-12-V_{M_2} - H) \right],
\nonumber\\
a^p_{8}(M_1 M_2) & = & c_8 + {c_7 \over N_c}
\left( 1 - 6{C_F \alpha_s\over 4
\pi} \right) + {\alpha\over 9\pi N_c} P^{p,EW}_{M_2, 3},
\nonumber\\
a_9(M_1 M_2) & = & c_9 + {c_{10} \over N_c}
\left[ 1 + {C_F \alpha_s \over 4 \pi}
(V_{M_2} +  H) \right],
\nonumber\\
a^p_{10}(M_1 M_2) & = & c_{10} + {c_9\over N_c}
\left[ 1 + {C_F \alpha_s \over 4 \pi} (V_{M_2} +
H) \right] +
{\alpha \over 9 \pi N_c} P^{p,EW}_{M_2, 2},
\end{eqnarray}
where $C_F = (N^2_c-1)/2N_c$ and $N_c = 3$
is the number of colors.
The vertex, the hard gluon exchange with the
spectator and the penguin contributions, at
$\mu =m_b$, are:
\begin{eqnarray}
V_M &=& - 18 + \int^1_0 dx g(x)
\phi_M(x),
\nonumber\\
P^p_{M,2} &=& c_1 \left[
\frac{2}{3} + G_M(s_p) \right]
+ c_3 \left[
\frac{4}{3} + G_M(0) + G_M(1) \right]
\nonumber\\*
& & + \,  (c_4+c_6) \left[
(n_f-2) G_M(0) + G_M (s_c) + G_M(1) \right]
\nonumber \\*
& & - \, 2 c_{8g}^{eff} \int^1_0 {dx\over 1-x}\phi_M(x),
\nonumber\\
P^{p,EW}_{M,2}& =& (c_1 + N_c c_2) \left[
\frac{2}{3} + G_M(s_p) \right] -
3 c_{7\gamma}^{eff} \int^1_0 {dx\over 1-x} \phi_M(x),
\nonumber\\
P^p_{M,3} &=& c_1 \left[ \frac{2}{3} + \hat{G}_M(s_p)
\right] + c_3 \left[ \frac{4}{3} + \hat{G}_M(0) +
\hat{G}_M(1) \right]
\nonumber\\*
& & + \, (c_4+c_6) \left[ (n_f-2) \hat{G}_M(0) +
\hat{G}_M(s_c) + \hat{G}_M(1) \right]
- 2 c_{8g}^{eff},
\nonumber\\
P^{p,EW}_{M,3} &=& (c_1 + N_c c_2)
\left[ \frac{2}{3} + \hat{G}_M(s_p) \right] -
3 c_{7\gamma}^{eff},
\nonumber\\
H &=& \frac{4 \pi^2}{N_c} \frac{f_B f_{M_1}}{m_B^2
F^{B\to M_1}_0(0)}
\nonumber \\*
& & \times
\int^1_0 {d \xi \over \xi} \phi_B(\xi)
\int^1_0 {dx\over \bar{x}} \phi_{M_2}(x)
\int^1_0 {dy \over \bar{y}} \left[
\phi_{M_1}(y) + \frac{2 \mu_{M_1}}{m_b}
\frac{\bar{x}}{x} \phi^p_{M_1}(y) \right] ,
\; \; \; \; \; \;
\end{eqnarray}
where $\bar{x}= 1-x$, $\bar{y} = 1-y$
and the parameter $2\mu_M/m_b$ coincides with $r_\chi$.
The functions $g(x)$, $G_M(x)$
and $\hat G_M(x)$ are given by
\begin{eqnarray}
g(x) & = & 3 \left( {1-2x\over 1-x}\ln x - i\pi \right)
\nonumber \\*
& & +
\left[ 2 {\rm Li}_2(x) -\ln^2x + {2\ln x\over 1-x} -
(3+2i\pi)\ln x -(x \to \bar{x} ) \right],
\nonumber\\
G(s,x) & = & 4 \int^1_0 du \, u
\bar{u} \ln[s-u \bar{u} x]
\nonumber \\*
& = & - \frac{10}{9} + \frac{2}{3} \ln s - \frac{8s}{3x}
+ \frac{4}{3} \left( 1 + \frac{2s}{x} \right)
\sqrt{\frac{4s}{x} -1} \arctan \frac{1}{\sqrt{\frac{4s}{x}
-1}} \, ,
\nonumber \\
G_M(s) & = & \int^1_0 dx \, G(s-i\epsilon,
\bar{x}) \phi_M(x),
\nonumber\\
\hat G_M(s) & = & \int^1_0 dx \,
G(s-i\epsilon, \bar{x}) \phi^p_M(x),
\end{eqnarray}
where $s_i  = m_i^2/m_b^2$ are the mass ratios
for the quarks involved in the penguin diagrams,
namely $s_u = s_d = s_s =0$ and $s_c = (1.3/4.2)^2$.

As it can be noticed,
except for the hard contribution where the
wave functions for both $M_1$ and $M_2$ are involved,
the coefficients $a_i$ are
different for the $X$ and $Y$ final states, since
they depend on the twist-2 and twist-3 wave functions
of the $M_2$ meson.
Thus, the twist-2 distribution amplitude $\phi_K(x)$ has
the following expansion in Gegenbauer polynomials
[12, 16]
\begin{equation}
\phi_K(x) = 6x(1-x)[ 1+ \alpha_1^K C^{(3/2)}_1(2x-1) +
\alpha_2^K C^{3/2}_2(2x-1) + ...],
\end{equation}
with $C^{3/2}_1(u) = 3 u$,
$C^{3/2}_2(u) = (3/2)(5u^2-1)$,
$\alpha_1^K = 0.3 \pm 0.3$, and
$\alpha_2^K = 0.1 \pm 0.3$. The
corresponding twist-3 amplitude, $\phi^p_K$, is 1.

The physical states $\eta$ and
$\eta^{\prime}$ are mixtures of SU(3)-singlet
and octet components $\eta_0$ and $\eta_8$ and
therefore the corresponding decay constants, in the
two-angle mixing formalism, are given by
\begin{eqnarray}
f^u_{\eta^{\prime}} & = &
\frac{f_8}{\sqrt{6}} \sin \theta_8 +
\frac{f_0}{\sqrt{3}} \cos \theta_0 \; ,
\nonumber \\
f^s_{\eta^{\prime}} & = &
- \, 2 \frac{f_8}{\sqrt{6}} \sin \theta_8 +
\frac{f_0}{\sqrt{3}} \cos \theta_0 \; ,
\end{eqnarray}
with $\theta_8 = -22.2^o$, $\theta_0 = -9.1^o$,
$f_8 = 168$ MeV, and $f_0 =157$ MeV [17]. These lead to
$f^u_{\eta^{\prime}} = 63.5$ MeV,
$f^s_{\eta^{\prime}} = 141$ MeV and to the relevant
form factor for the $B \to \eta^{\prime}$ transition
\begin{equation}
F_0^{B \to \eta^{\prime}} =
F_0^{\pi} \left( \frac{\sin \theta_8}{\sqrt{6}}
+ \frac{\cos \theta_0}{\sqrt{3}} \right) = 0.137
\end{equation}

Even though the $\eta^{\prime}$ flavor singlet meson
has a gluonic content which could bring a
contribution to the wave function, this is
supposed to be small [18] and therefore we
employ, in the calculation of
$V_{\eta^{\prime}}$, $P^p_{\eta^{\prime},2}$
and $P^{p,EW}_{\eta^{\prime},2}$ in $a_i(Y)$,
only the leading twist-2 distribution amplitude
\begin{equation}
\phi_{\eta^{\prime}} = 6 x \bar{x} \, .
\end{equation}
Also, since the twist-3 quark-antiquark distribution
amplitude do not contribute, due to the chirality
conservation, the penguin parts in $a_6^p(Y)$ and
$a_8^p(Y)$ are missing.
As for the $B$ meson wave function,
we shall work with a strongly peaked one, around
$z_0 = \lambda_B/m_B \approx 0.066 \pm 0.029$,
for $\lambda_B =0.35 \pm 0.15$ GeV.

Putting everything together, we get, within
the SM improved factorization approach [12], the
numerical value $BR_{SM}(B \to \eta^{\prime} K ) =
3.65 \times 10^{-5}$, which although is in accordance
with other theoretical estimations [5, 15, 17],
yet lay below the experimental data [1-4].
Hence, in spite of the ^^ ^^ conservative''
prediction that the conventional mechanism
should be the dominant one, it has been
getting clear that new contributions are needed
in order to account for the existent data.

\section{Spectator Hard-Scattering Mechanism}

It has been considered that the spectator hard-scattering
mechanism (SHSM), depicted in Figure 1, is a reliable framework
for this process, which significantly increases the value of $BR(B
\to \eta^{\prime} K)$ [5, 13]. Following this idea, let us write
down the corresponding di-gluon exchange amplitude for the $b$
quark decaying into an $s$ quark and a hard gluon
\begin{eqnarray}
A_{hs} & = & - \, i \, C_F \,
g_s^3 \frac{f_B}{2 \sqrt{6}}
\frac{f_K}{2 \sqrt{6}}
\int dz \, dy \, \phi_B(z) \phi_K(y)
\nonumber \\*
& & \times \, {\rm Tr} \left[ \gamma_5 \rlap{/}{P_k} \Gamma_{\mu}
( \rlap{/}{P_B} + m_B) \gamma_5 \gamma_{\nu} \right]
\frac{\varepsilon^{\mu \nu \alpha \beta} Q_{1 \alpha}
Q_{2 \beta}}{Q_1^2 Q_2^2} \,
F_{\eta^{\prime} g^* g^*}
(Q_1^2, Q_2^2, m_{\eta^{\prime}}^2) \; \; \; \;
\end{eqnarray}
in terms of the effective $b \to sg$ vertex [19]
\begin{equation}
\Gamma_{\mu}^a = \frac{G_F}{\sqrt{2}} \,
\frac{g_s}{4 \pi^2} V^*_{ps} V_{pb} \, t^a
\left[ F_1^p \left( Q_1^2 \gamma_{\mu}
- Q_{1 \mu} Q_1 \right) L
- F_2^p i \sigma_{\mu \nu} Q_1^{\nu} m_b R \right]
\end{equation}
and the transition form factor [6]
\begin{equation}
< g_a^* g_b^* | \eta^{\prime}> = - \, i \,
\delta_{ab} \varepsilon^{\mu \nu \alpha \beta}
\varepsilon^{a*}_{\mu} \varepsilon^{b*}_{\nu}
Q_{1 \alpha} Q_{2 \beta} F_{\eta^{\prime} g^* g^*}
(Q_1^2, Q_2^2, m_{\eta^{\prime}}^2)
\end{equation}
The quark contribution to the
$\eta^{\prime} g^* g^*$ vertex
\begin{equation}
F_{\eta^{\prime} g^* g^*}
(Q_1^2, Q_2^2, m_{\eta^{\prime}}^2) =
4 \pi \alpha_s \frac{1}{2N_c} \sum_{q=u,d,s}
f^q_{\eta^{\prime}} \, F(y,a) \, ,
\end{equation}
with
\begin{equation}
F(y,a) =
\int_0^1 dx \; \frac{\phi_{\eta^{\prime}}(x)}{\bar{x}
Q_1^2 + x Q_2^2 -x \bar{x} m_{\eta^{\prime}}^2 + i
\varepsilon} + ( x \leftrightarrow \bar{x} )
\; , \; \; a^2=m_{\eta^{\prime}}^2/m_B^2 \, ,
\end{equation}
will play an important role in the evaluation of the
amplitude $A_{hs}$.
Performing the calculations in (9), we come to the
following expression of the hard scattering amplitude:
\begin{eqnarray}
A_{hs} & = &
- \, 2 \, i \, \frac{G_F}{\sqrt{2}} V_{ps}^*
V_{pb} \frac{\alpha_s^2}{N_c^3} f_B f_K
(2f^u_{\eta^{\prime}} + f^s_{\eta^{\prime}})
\int_0^1 dz \, \phi_B(z) \int_0^1 dy \phi_K(y)
\nonumber \\*
& &
\times \left[ F_1^p Q_1^2 \left( (P_B \cdot Q_1)
(P_K \cdot Q_2) - (P_K \cdot Q_1)(P_B \cdot Q_2)
\right) + \right.
\nonumber \\*
& & + \, \left.
F_2^p m_B m_b \left( (P_K \cdot Q_2)
Q_1^2 - (P_K \cdot Q_1)(Q_1 \cdot Q_2) \right)
\right] \frac{F(y,a)}{Q_1^2 Q_2^2}
\end{eqnarray}
With the gluon momenta
\begin{equation}
Q_1=\bar{z}
P_B - \bar{y}P_K \; , \; \; Q_2 = zP_B -yP_K \; ,
\end{equation}
and neglecting, for the moment, both
$m_{\eta^{\prime}}^2$ and $m_K^2$,
the amplitude (14) becomes
\begin{eqnarray}
A_{hs} & = &
i \, \frac{G_F}{\sqrt{2}} \, V_{ps}^* V_{pb}
\, \frac{\alpha_s^2}{2 N_c^3}
f_B f_K (2f^u_{\eta^{\prime}} + f^s_{\eta^{\prime}})
\, \frac{1}{z_0}
\nonumber \\*
& & \times \,
\int_0^1 \phi_K(y) \left[ m_B^2 F_1^p +
m_B m_b \frac{F_2^p}{y-z_0} \right] F(y,a)
\end{eqnarray}
where, for the dominant contribution
coming from the insertion of the $O_1^{u,c}$
and the magnetic-penguin $O_{8g}$ operators,
one has [13]
\begin{equation}
F_1^p = c_1 \left[ \frac{2}{3} + G[s_p,
(1-z_0)(y-z_0)] \right] \, , \; \;
F_2^p = - 2 c_{8g}
\end{equation}
In what it concerns the
$F(y,a)$ function, which is an essential
input in the calculations,
it can be first written as
\begin{equation}
F(y,a) \, = \,
4 \int_0^1 dx \,  \frac{6x \bar{x}
\; (Q_1^2+Q_2^2-2 x \bar{x}
m_{\eta^{\prime}}^2)}{\left[ Q_1^2+Q_2^2-2x
\bar{x} m_{\eta^{\prime}}^2
\right]^2 - \left[
(x - \bar{x}) (Q_1^2 - Q_2^2) \right]^2}
\end{equation}
and it comes, after algebraic computations,
to the following form
\begin{eqnarray}
F(y,a) & = &
- \, \frac{12}{m_{\eta^{\prime}}^2} \left[
1 \, - \, \frac{Q_1^2-Q_2^2}{2 m_{\eta^{\prime}}^2}
\log \left| \frac{Q_1^2}{Q_2^2} \right| +
\, \frac{(Q_1^2-Q_2^2)^2 -
m_{\eta^{\prime}}^2 ( Q_1^2+Q_2^2)}{2
m_{\eta^{\prime}}^2 \sqrt{p^4-4 Q_1^2 Q_2^2}}
\right. \nonumber \\*
& & \times \left. \log \left|
1 + 2 \, \frac{\sqrt{p^4-4Q_1^2Q_2^2}}{p^2
- \sqrt{p^4-4Q_1^2Q_2^2}} \right|
\right],
\end{eqnarray}
where we have introduced the notation
$p^2 = Q_1^2 + Q_2^2 -m_{\eta^{\prime}}^2$.
The logarithmic nature of the $F(y,a)$ function
makes it very sensitive to the values
of $Q_1^2$, $Q_2^2$, $m_{\eta^{\prime}}^2$.
We recommend [6] for a detailed discussion
of the $\eta^{\prime} g^* g^*$ vertex in the case
of arbitrary gluon virtualities in the
time-like, $Q_1^2>0$, $Q_2^2 >0$,
$p^4-4Q_1^2Q_2^2 >0$, and space-like,
$Q_1^2<0$, $Q_2^2 <0$,
$p^4-4Q_1^2Q_2^2 <0$, regions.

Now, using
\begin{equation}
Q_1^2 \approx \bar{z} \left[(y-z)m_B^2+\bar{y}
m_{\eta^{\prime}}^2 \right] \, , \; \;
Q_2^2 \approx z \left[ -(y-z)m_B^2+ y
m_{\eta^{\prime}}^2 \right] ,
\end{equation}
where we have  neglected $m_K^2$,
the dominant term in (19) is:
\begin{eqnarray}
F(y,a) & \approx &
- \, \frac{12}{m_{\eta^{\prime}}^2} \left[
1 \, - \, \frac{1}{2} \left[ \frac{y-z}{a^2}
+ (1-y-z) \right] \log \left| \frac{a^2+y-z}{z(z-y)}
\right| \right.
\nonumber \\*
& & + \, \left.
\frac{(y-z)a^2+(y-z)^2}{2a^2 \; |y-z|} \log
\left| \frac{y(1-a^2)-z+|y-z|}{y(1-a^2)
-z-|y-z|} \right| \right]
\end{eqnarray}

On the other hand, by comparing the expressions in (20), it
clearly results that we are in the limit where $|Q_1^2| \gg
|Q_2^2|$. So, the function $F(y,a)$ can be computed in this
approximation and it simply yields
\begin{equation}
F (y,a) \, = \,
- \, \frac{12}{m_{\eta^{\prime}}^2}
\left[ 1 + \left( \frac{y-z_0}{a^2} + \bar{y} \right)
\log \left| 1-
\frac{1}{\frac{y-z_0}{a^2}+ \bar{y}} \right| \right]
\end{equation}
As it can be seen from (20), the term
$(y-z_0)/a^2 + \bar{y} = Q_1^2/m_{\eta^{\prime}}^2$
takes a whole range of values,
from $-0.87$ to $26.5$, as $Q_1^2$ goes from
the space-like to the time-like regions.
Consequently, a logarithmic singularity develops as
$y \to z_0/(1-a^2)$, i.e. for
$Q_1^2 \to m_{\eta^{\prime}}^2$.
Inspecting (16), we also notice the pole
at $y = z_0$ in the $F_2^p$ contribution.
In addition, while $G[s_p,(1-z_0)(y-z_0)]$ is
divergence free for all $s>0$, the
$G[0,(1-z_0)(y-z_0)]$ gets a logarithimic singularity
at $y=z_0$.
Hence, in the course of numerically evaluating the
scattering contribution, one must be
careful about dealing with these combined
singularities in the convolution (16).

As in the case of other hard-scattering theoretical estimations
[5, 13], the amplitude of this contribution contains, as main
uncertainty, the peaking position, $z_0$, in the $B$ meson
distribution function and accordingly, the branching ratio is
extremely sensitive to it. For $z_0 \in [0.063 , \, 0.068]$ and
the average value $\alpha_s(Q_1^2)=0.28$, the total branching
ratio, including besides the improved factorization approach, the
spectator hard-scattering mechanism with the vertex function (22),
is in the range from $BR(B \to \eta^{\prime} K) = 6.58 \times
10^{-5}$, for $z_0 =0.063$, to $BR(B \to \eta^{\prime} K) = 5.8
\times 10^{-5}$, for $z_0 =0.068$.

Comparing these results with the experimental data [$1-3$], we
notice that they are still below the lowest limit. An alternative
way which increases the $BR$ and avoids the uncertainties coming
from the combined singularities in the convolution (16), would
presumably look more reliable.

\section{SUSY Gluonic Dipole Contribution}

Employing the Minimal Supersymmetric
Standard Model (MSSM), we shall add to the effective SM
Hamiltonian (1), the SUSY contribution
\begin{equation}
H^{SUSY} \, = \,  - \, i \, {G_F \over \sqrt{2}}
(V_{ub} V_{us}^* +V_{cb} V_{cs}^*)
\left( c_{8g}^{SUSY} O_{8g} + c_{7 \gamma}^{SUSY}
O_{7 \gamma} \right) ,
\end{equation}
expressed in terms of the usual gluon
and photon operators:
\begin{eqnarray}
O_{8g} & = & \frac{g_s}{8 \pi^2} \, m_b
\bar{s} \sigma_{\mu \nu} (1+\gamma_5 ) G^{\mu \nu} b \, ,
\nonumber \\
O_{7 \gamma} & = & \frac{e}{8 \pi^2} \, m_b \bar{s}
\sigma_{\mu \nu} (1+\gamma_5 ) F^{\mu \nu} b \, ,
\end{eqnarray}
and of the Wilson coefficients [10, 20]
\begin{eqnarray}
c_{8g}^{SUSY} (M_{SUSY}) & = & - \, \frac{\sqrt{2} \pi
\alpha_s}{G_F (V_{ub} V_{us}^* +
V_{cb} V_{cs}^*) m_{\tilde{g}}^2} \, \delta^{bs}_{LR} \,
\frac{m_{\tilde{g}}}{m_b} \, G_0(x) \, ,
\nonumber \\
c_{7 \gamma}^{SUSY} (M_{SUSY})  & = & - \,
\frac{\sqrt{2} \pi \alpha_s}{G_F (V_{ub} V_{us}^* +
V_{cb} V_{cs}^*) m_{\tilde{g}}^2} \, \delta^{bs}_{LR} \,
\frac{m_{\tilde{g}}}{m_b} \, F_0(x) \, ,
\end{eqnarray}
where
\begin{eqnarray}
G_0 (x) & = &
\frac{x}{3(1-x)^4} \,
\left[ 22-20x-2x^2+16 x \ln(x) -x^2 \ln (x)
+ 9 \ln (x) \right] ,
\nonumber \\
F_0 (x) & = & - \;
\frac{4x}{9(1-x)^4} \,
\left[ 1+4x-5x^2+4 x \ln(x) + 2 x^2 \ln (x) \right]
\end{eqnarray}
In the above expressions, $x=m_{\tilde{g}}^2 / m_{\tilde{q}}^2$,
with $m_{\tilde{g}}$ being the gluino mass and $m_{\tilde{q}}$ an
average squark mass, while the factor $\delta^{bs} = \Delta^{bs}/
m_{\tilde{q}}^2$, where $\Delta^{bs}$ are the off-diagonal terms
in the sfermion mass matrices, comes from the expansion of the
squark propagator in terms of $\delta$, for $\Delta \ll
m_{\tilde{q}}^2$. In principle, the dimensionless quantities
$\delta^{bs}$, measuring the size of flavor changing interaction
for the $\tilde{s} \tilde{b}$ mixing, are present in all the SUSY
corrections to the Wilson coefficients in (1) and they are of four
types, depending on the $L$ or $R$ helicity of the fermionic
partners. In the followings, we focus on the $\delta^{bs}_{LR}$
insertions because only the SUSY Wilson coefficients (25), being
proportional to the large factor $m_{\tilde{g}} /m_b$, are going
to make an important contribution, even for small values of
$\delta$.

In (3), we replace the Wilson coefficients $c_{8g}^{eff}$ and
$c_{7 \gamma}^{eff}$, by the total quantities
\begin{equation}
c_{8g}^{total} [x, \delta] \, = \,
c_{8g}^{eff} + c_{8g}^{SUSY} (m_b) \; ,
\; \;
c_{7 \gamma}^{total} [x, \delta] \, = \,
c_{7 \gamma}^{eff} + c_{7 \gamma}^{SUSY} (m_b) \; ,
\end{equation}
where
$c^{SUSY} (m_b)$ have been evolved from
$M_{SUSY} = m_{\tilde{g}}$ down to the
$\mu =m_b$ scale, using the relations [10, 19]
\begin{eqnarray}
c_{8g}^{SUSY} (m_b) & = &
\eta c_{8g}^{SUSY}(m_{\tilde{g}} ) \, ,
\nonumber \\
c_{7 \gamma}^{SUSY} (m_b) & = & \eta^2
c_{7 \gamma }^{SUSY}(m_{\tilde{g}} ) + \frac{8}{3}
(\eta - \eta^2) c_{8g}^{SUSY}(m_{\tilde{g}} ) \, ,
\end{eqnarray}
with
\begin{equation}
\eta \, = \,
\left( \alpha_s(m_{\tilde{g}})/
\alpha_s(m_t) \right)^{2/21}
\left( \alpha_s(m_t)/
\alpha_s(m_b) \right)^{2/23}
\end{equation}

We choose for $m_{\tilde{q}}$ the value
$m_{\tilde{q}} = 500$ GeV and write
$m_{\tilde{g}}$ as $m_{\tilde{g}} = \sqrt{x} \,
m_{\tilde{q}}$ and $\delta^{bs}_{LR} \equiv \rho e^{i \varphi}$.
As the total branching ratio
can be expressed in terms of three free parameters:
$x , \, \rho , \, \varphi$,
one is able to plot the $BR^{total}$,
in units of $10^{-5}$, as a function of $(\rho , \varphi)$,
for different values of $x$.
By inspecting the 3D plots displayed in Figure 2, for
$x=0.3$ (the upper) and $x=1$ (the lower surface), we notice
that the SUSY contributions (25) to the Wilson coefficients
have significantly increased the SM value,
$BR_{SM} =3.65 \times 10^{-5}$, represented by the horizontal plane.
Using the experimental data, one is able now to determine
the $\delta^{bs}_{LR}$ complex values, for each $x$.

Let us take, for example, $x=1$, pointing out that the same
discussion can be performed for any $x$-value. For $\rho =0.005$,
the $BR^{total}$ is increasing from $5.1 \times 10^{-5}$, for
$\varphi \approx \pm \pi /3$, to the maximum value $BR^{total}=
6.24 \times 10^{-5}$, for $\varphi =0$. As $\rho$ goes to bigger
values, we find a better agreement with the large experimental
data. For $\rho =0.01$, the data can be accommodated for $\varphi
\approx -\pi /4$, while, for $\rho=0.02$, one has to impose
$\varphi \approx - \, 8 \pi /15$.

\section{Concluding Remarks}

At first, we have analyzed the $B^- \to \eta^{\prime} K$ decay and
computed its branching ratio using the improved factorization
method developed by Beneke {\it al.} [12]. Since the obtained
result, $BR_{SM} = 3.65 \times 10^{-5}$, is much below the
experimental data, [1-4], we have have added new contributions.

In this respect, the so-called spectator hard-scattering
mechanism, which is depicted in Figure 1, has allowed us to
compute the amplitude in terms of the effective $b \to sg$ vertex
and the transition form factor (11) which contains the quark
contribution to the $\eta^{\prime} g^* g^*$, (21), as an essential
input. The total $BR$ has, as a main uncertainty, the peaking
position in the $B$ meson wave function, $z_0 = \lambda_B/m_B$,
with $\lambda_B = 0.35 \pm 0.15$ GeV. Even the results are closer
to the experimental data, we point out the combined singularities
in the amplitude convolution (16) which must be treated carefully.

Secondly, we extend the SM to the MSSM and add SUSY contributions
to the Wilson coefficients $c^{eff}_{8g}$ and $c^{eff}_{7
\gamma}$. The total $BR$ is expressed in terms of the parameters
$x=m_{\tilde{g}}^2/m_{\tilde{q}}^2$, and $\delta^{bs}_{LR} = \rho
e^{i \varphi}$ whose contribution turns out to be important, even
for very small values of $\rho$. Finally, by inspecting the
3D-graphics (see Figure 2), representing the $BR^{total}$ for
$x=0.3$ (the upper surface) and $x=1$ (the lower surface), one is
able to find numerical values for $\rho$ and $\varphi$ that can
account for the experimental data or other theoretical predictions
[21].

\begin{flushleft}
\begin{Large}
{\bf Acknowledgments}
\end{Large}
\end{flushleft}

The authors wish to acknowledge the
kind hospitality and fertile environment
of the University of Oregon
where this work has been carried out.
Professor N.G. Deshpande's inciting suggestions and
constant support are highly regarded.
M.A.D. thanks the U.S. Department of State, the Council
for International Exchange of Scholars (C.I.E.S.)
and the Romanian-U.S. Fulbright Commission for
sponsoring her participation in the
Exchange Visitor Program no. G-1-0005.

\newpage

\begin{center}
{\bf Figure Captions}
\end{center}
Fig.1. Feynman diagrams of the hard scattering mechanism for
$B^- \to \eta^{\prime} K^-$. The gluons are represented
by the dashed lines. \\
Fig.2. Total branching ratios (SM+SUSY) for $B^- \to \eta^{\prime}
K^-$, in units of $10^{-5}$, as functions of $(\rho , \varphi)$,
for $x=0.3$ (the upper plot) and $x=1$ (the lower plot), compared
to the SM estimation represented by the horizontal plane.

\newpage
\vspace*{2cm}

\begin{figure}
\includegraphics{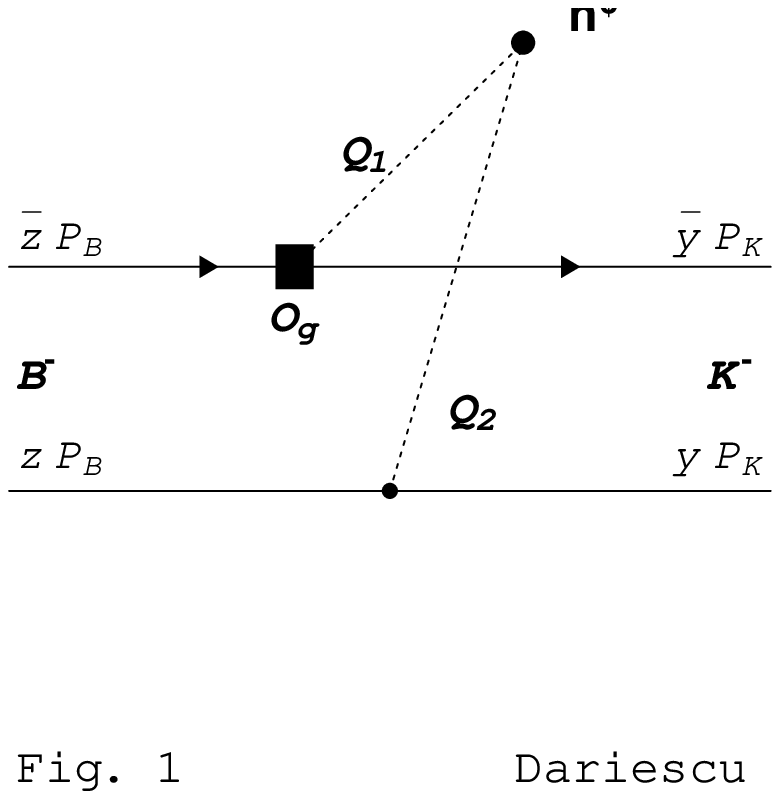}
\end{figure}

\begin{figure}
\includegraphics{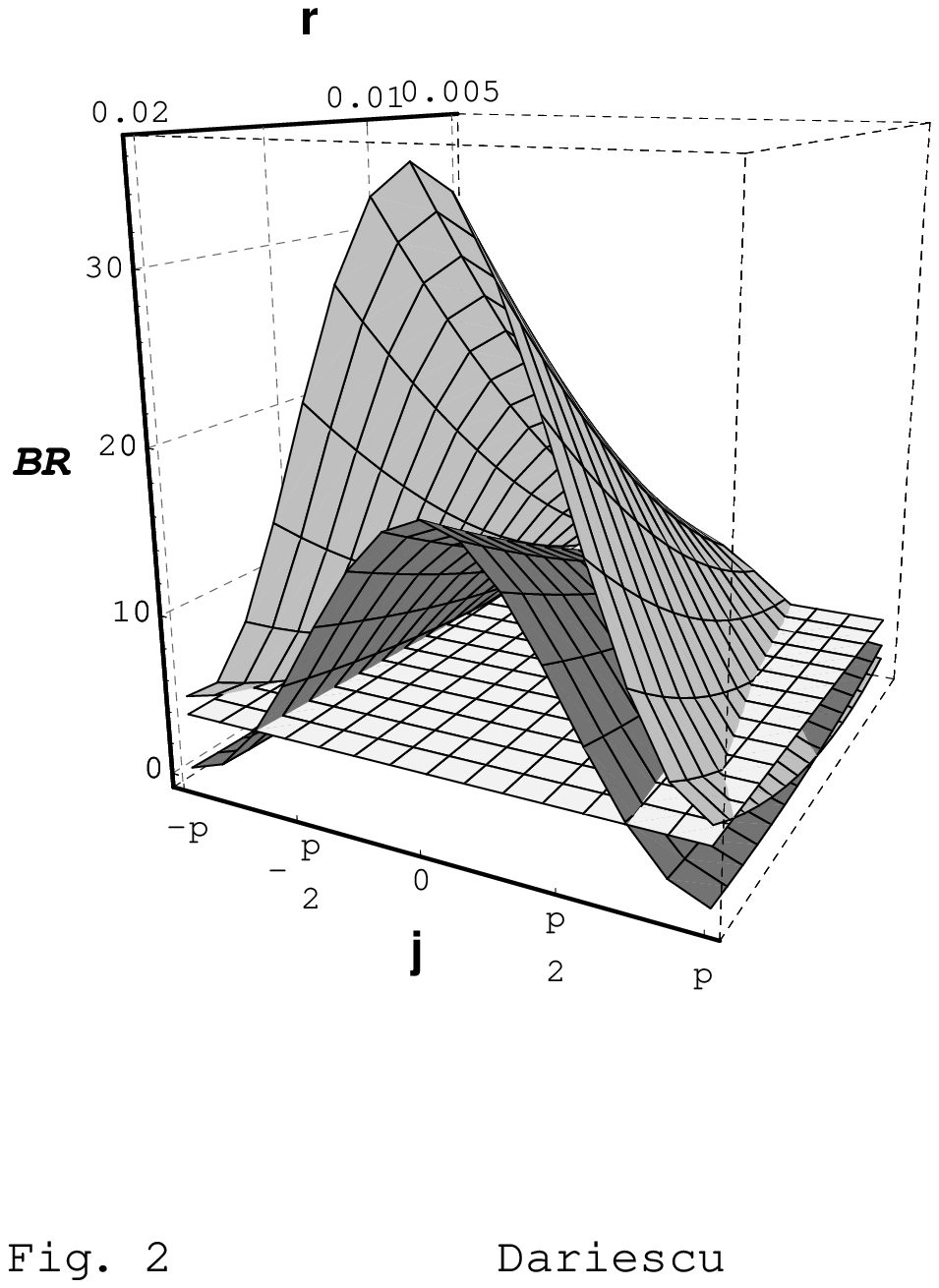}
\end{figure}

\end{document}